\documentclass[11pt]{iopart}
\usepackage{iopams,epstopdf,graphicx,url}
\expandafter\let\csname equation*\endcsname\relax
\expandafter\let\csname endequation*\endcsname\relax

\newcommand{\Tlib}{T_{\mbox{\scriptsize lib}}}
\newcommand{\Tobs}{T_{\mbox{\scriptsize obs}}}
\newcommand{\Tlbar}{\bar{T}_{\mbox{\scriptsize lib}}}
\newcommand{\Tobar}{\bar{T}_{\mbox{\scriptsize obs}}}

\begin{document}

\title{Expanded solar-system limits on violations of the equivalence principle}

\author{James Overduin,$^{1,2}$ Jack Mitcham$^1$ and Zoey Warecki$^1$}
\address{$^1$ Department of Physics, Astronomy and Geosciences, Towson University, Towson, MD 21252}
\address{$^2$ Department of Physics and Astronomy, Johns Hopkins University, Baltimore, MD 21218}
\ead{joverduin@towson.edu}

\begin{abstract}
Most attempts to unify general relativity with the standard model of particle physics predict violations of the equivalence principle associated in some way with the composition of the test masses.
We test this idea by using observational uncertainties in the positions and motions of solar-system bodies to set upper limits on the relative difference $\Delta$ between gravitational and inertial mass for each body.
For suitable pairs of objects, it is possible to constrain three different linear combinations of $\Delta$ using Kepler's third law, the migration of stable Lagrange points, and orbital polarization (the Nordtvedt effect).
Limits of order $10^{-10}-10^{-6}$ on $\Delta$ for individual bodies can then be derived from planetary and lunar ephemerides, Cassini observations of the Saturn system, and observations of Jupiter's Trojan asteroids as well as recently discovered Trojan companions around the Earth, Mars, Neptune, and Saturnian moons.
These results can be combined with models for elemental abundances in each body to test for composition-dependent violations of the universality of free fall in the solar system.
The resulting limits are weaker than those from laboratory experiments, but span a larger volume in composition space.
\end{abstract}

\maketitle

\section{Introduction} \label{sec:intro}

The foundation of general relativity is the equivalence principle (EP), the weak version of which states that inertial mass and gravitational mass are identical.
Attempts to unify general relativity with the standard model of particle physics generically predict the existence of new fields with gravitational-strength couplings to existing standard-model fields.
But these couplings are not universal, in contrast to the coupling between standard-model fields and the metric or spin-two graviton field of general relativity.
Hence they introduce differences in the accelerations of test bodies in the same gravitational field, violating the EP.
Such violations can be quantified with the parameter $\Delta$, defined by
\begin{equation}
m_g/m_i \equiv 1 + \Delta
\label{defnDelta}
\end{equation}
where $m_g$ is the gravitational mass and $m_i$ is the inertial mass.
In some theories $\Delta$ is associated with gravitational self-energy $U$, so that $\Delta = \eta U$, and experimental constraints on $\Delta$ effectively translate into upper limits on a universal constant $\eta$ \cite{Nordtvedt1968,Orellana1993}.
In others, the value of $\Delta$ may in principle vary from object to object depending on composition or other factors \cite{Overduin2000,Blaser2003,Overduin2012}.

Three main approaches have been taken in testing the EP.
The oldest, and  in principle the simplest, is to drop two objects with different properties in the same gravitational field (as may have been done by Galileo using a lead musketball and iron cannonball at Pisa, and was definitely done by some of his contemporaries, such as Simon Stevin using lead balls in Delft).
For these early investigators, the main property of interest was test-body mass.
Nowadays we might be more interested in composition.
All energy gravitates, but all forms of energy may not couple in the same way to the new fields predicted by modern unified theories.
In ``runaway dilaton'' versions of string theory copper and beryllium 
%(or platinum and titanium) 
fall at different rates due to factors involving differences in electromagnetic binding energy \cite{Damour1994,Damour2002}.
Modest extensions of the standard model with a single minimally-coupled scalar field predict EP violations for test materials such as aluminum and beryllium due primarily to couplings between this field and gluons, the gauge fields of quantum chromodynamics \cite{Carroll1998,Chen2005}.
Other theories involving ``little strings'' \cite{Antoniadis2001}, time-varying fundamental ``constants'' \cite{Dvali2002,Wetterich2003}, ``chameleon fields'' \cite{Khoury2004,Mota2007,Capozziello2008} and generic violations of Lorentz symmetry \cite{Kostelecky2009,Kostelecky2010} have similar consequences.

The scale of predicted EP violations in these theories is however very small, of order $10^{-12}$ or less.
To detect them, modern versions of Galileo's drop-tower experiment must be carried out in space, where macroscopic test bodies can fall continuously in a disturbance-free environment over many orbits around the earth.
One such experiment, MicroSCOPE, is designed to measure the relative accelerations of two pairs of test masses composed of platinum and titanium alloys with a sensitivity of $10^{-15}$ \cite{Touboul2009}.
Another, the Satellite Test of the Equivalence Principle (STEP), aims to reach a sensitivity of $10^{-18}$ through the use of superconducting accelerometers, and to monitor four pairs of beryllium, niobium and platinum-iridium test masses \cite{Sumner2007,Overduin2012}.
Alternatively, it may be possible to reach comparable levels of precision with {\em microscopic\/} test particles using atom interferometry in the laboratory \cite{Dimopoulos2008} or in space, as in the proposed Space-Time Explorer and QUantum Equivalence Principle Space Test (STE-QUEST) \cite{Galoul2013}.
At present such techniques are limited to different isotopic pairs of the same element, such as rubidium-85 and 87, but experiments involving lithium and cesium are envisioned for the future \cite{Kim2010}.

The second historical approach to tests of equivalence makes use of sensitive torsion balances to compare the accelerations of different objects in what is effectively a horizontal component of the gravitational field of the Earth, Sun or Milky Way.
Pioneered by E\"{o}tv\"{o}s in the nineteenth century, this technique has produced the strongest current constraint on EP violation, limiting any difference in acceleration of beryllium and titanium in the field of the Earth to less than $(0.3 \pm 1.8) \times 10^{-13}$ \cite{Schlamminger2008}.

The third EP testing strategy, and the one that is the focus of this work, follows from Newton's realization that nature provides a ``free'' way to test our theories of gravity in the form of the continuously falling moons and planets of the solar system \cite{Harper2002}.
Celestial EP tests open up regions of parameter space inaccessible to terrestrial experiment (for example, comparing the accelerations of what are effectively a ball of hydrogen and a ball of rock and metal).
However they are not generally as sensitive as torsion-balance experiments or proposed free-fall tests in space.
The major exception so far involves the phenomenon of orbital polarization (the Nordtvedt effect), whereby the elliptical orbit of one body around a second becomes gradually aligned along the direction to a third, introducing anomalous variations in distance between the first two bodies.
Laser ranging using retroreflectors left on the moon by Apollo astronauts has made it possible to limit any such difference in accelerations of the Earth and Moon toward the Sun to less than $(-1.0 \pm 1.4) \times 10^{-13}$, comparable to the constraint from torsion balances \cite{Williams2004}.
Ranging to Mars may someday produce comparable results \cite{Anderson1996}.

Violations of the EP by solar system bodies also reveal themselves in modifications of Kepler's third law and migrations of stable Lagrange points \cite{Nordtvedt1968}.
These effects do not generally produce individual limits as strong as those from orbital polarization \cite{Orellana1993}.
However, they constrain two linearly independent combinations of $\Delta$ parameters, and are therefore particularly useful in testing theories for which the value of $\Delta$ can differ from object to object.
This method has been applied, for example, to put the strongest current limits on extensions of general relativity to higher dimensions, where static, spherically symmetric objects like stars or planets are models by generalizations of the Schwarzschild metric known as {\em solitons\/} \cite{Overduin2000}.
Our goal in this paper is to extend and strengthen this way of testing the EP using updated ephemerides and considering more objects, including additional Jupiter Trojans as well as companions of the Earth, Mars, Neptune and Saturn's moons Tethys and Dione.

We investigate Kepler's third law, the migration of Lagrange points and orbital polarization in Sections~\ref{sec:kepler}, \ref{sec:lagrange} and \ref{sec:nordtvedt} respectively.
Limits on individual solar-system bodies are derived in Sec.~\ref{sec:individual}.
In Sec.~\ref{sec:elemental} we discuss applications of these results, and use models of the compositions of these bodies to derive limits on EP violation by individual constituent elements, assuming that a single element dominates in each case.
We conclude with a summary and discussion in Sec.~\ref{sec:discussion}.

\section{Modified Kepler's third law} \label{sec:kepler}

If two bodies with gravitational masses $m_1,m_2$ both violate the EP according to Eq.~(\ref{defnDelta}) then extra terms appear in Kepler's third law \cite{Nordtvedt1968,Overduin2000}
\begin{equation}
G(m_1+m_2+m_2\Delta_1+m_1\Delta_2) = \omega^2a^3 ,
\label{Kepler0}
\end{equation}
where $\Delta_1$ and $\Delta_2$ are the EP violating parameters for $m_1$  and $m_2$ respectively.

The common part of $\Delta_1$ and $\Delta_2$ can be absorbed into a rescaled gravitational constant, as can be seen by rewriting Eq.~(\ref{Kepler0}) in the form
\begin{equation}
G(1+\Delta_1)(m_1+m_2)+Gm_1(\Delta_2- \Delta_1) = \omega^2a^3 .
\label{Kepler1}
\end{equation}
There are two modifications of Kepler's third law here: a rescaling of $G$ in the first term, and a completely new second term, which depends only on the difference $\Delta_2-\Delta_1$.
This latter term is a clear manifestation of EP violation in the system.

In practice, $G$ is avoided in celestial mechanics, since it is known only to about a part in $10^4$ \cite{Fixler2007}.
Since $G$ always appears together with a mass, it is common to work instead with $Gm_{\odot}=k^2A^3$ where $m_{\odot}$ is the mass of the Sun, $A$ is the length of the astronomical unit (AU) and $k$ is a defined constant (the Gaussian constant).
The value of $A$ (or equivalently, of $Gm_{\odot}$) can then be determined by fitting statistically to the entire history of observational data for all systems involving the sun.
Currently $A=149\,597\,870\,000$~m with $\delta A=\pm3$~m, an uncertainty of two parts in $10^{11}$ \cite{AstroAlmanac2012}.
We rewrite Eq.~(\ref{Kepler1}) to make better contact with observation as
\begin{equation}
\left(\!\frac{m_\odot}{m_1}\!\right) \! \left(\frac{\omega}{k}\right)^2 \!\! \left(\frac{a}{A}\right)^3 \!\! - \left(1 + \frac{1}{m_1/m_2}\right) = \frac{\Delta_1}{m_1/m_2} + \Delta_2 .
\label{Kepler2}
\end{equation}
In this form it is clear that $\Delta_1$ and $\Delta_2$ can be constrained experimentally, even in the special case where $\Delta_1=\Delta_2$.
We have set up the equation this way with the intent of applying it to systems where $m_1\gg m_2$ (i.e., where $m_2$ is in orbit around $m_1$).
While our knowledge of individual masses is subject to the same uncertainty as that in $G$, mass {\em ratios\/} can be measured with much higher precision using Kepler's third law.

\begin{table}[t!]
\footnotesize
\caption{Limits from Kepler's third law.}
\label{table-kepler}
\begin{tabular}{@{}llllllll}
\br
Pair ($m_{\mbox{\tiny 1}}$-$m_{\mbox{\tiny 2}}$) & $a \!$ (AU) & $\delta a \!$ (km) & $P \!$ (yr) & $\delta\omega \!$ (arcsec/cty) & $m_{\mbox{\tiny 1}}/m_{\mbox {\tiny 2}}$ & $\delta(m_{\mbox{\tiny 1}}/m_{\mbox {\tiny 2}})$ & $\epsilon_{\mbox {\tiny 1}}$\\
\mr
Sun-Mercury & 0.39 & 2 & 0.241 & 0.002 & $6.02\times10^6$ & $3.0\times10^2$ & $1\times10^{-7}$\\
Sun-Venus & 0.72 & 0.4 & 0.615 & 0.002 & $4.09\times10^5$ & $8.0\times10^{-3}$ & $1\times10^{-8}$\\
Sun-Earth & 1 & 0.006 & 1 & 0.002 & $3.33\times10^5$ & $7.0\times10^{-4}$ & $1\times10^{-10}$\\
Sun-Mars & 1.52 & 0.6 & 1.88 & 0.002 & $3.10\times10^6$ & $2.0\times10^{-2}$ & $8\times10^{-9}$\\
Sun-Jupiter & 5.20 & 20 & 11.9 & 0.2 & $1.05\times10^3$ & $1.7\times10^{-5}$ & $9\times10^{-8}$\\
Sun-Saturn & 9.53 & 0.6 & 29.5 & 0.2 & $3.50\times10^3$ & $1.0\times10^{-4}$ & $9\times10^{-8}$\\
Sun-Uranus & 19.2 & 400 & 84.3 & 0.2 & $2.29\times10^4$ & $3.0\times10^{-2}$ & $5\times10^{-7}$\\
Sun-Neptune & 30.1 & 2000 & 165 & 0.5 & $1.94\times10^4$ & $3.0\times10^{-2}$ & $2\times10^{-6}$\\
Earth-Moon & 384\,000$^{\ast}$ & 0.0012 & 27.3$^{\ast}$ & 0.01 & 81.3 & $3.0\times10^{-6}$ & $1\times10^{-8}$\\
Saturn-Tethys & 294\,000$^{\dag}$ & 0.02 & 191$^{\dag}$ & $4.2\times10^{-7}$$^{\dag}$ & $9.21\times10^5$ & 140 & $2\times10^{-7}$\\
Saturn-Dione & 377\,000$^{\dag}$ & 0.03 & 132$^{\dag}$ & $3.0\times10^{-7}$$^{\dag}$ & $5.19\times10^5$ & 18 & $2\times10^{-7}$\\
\br
\end{tabular}
$^\ast$For the Moon, $a$ is in km and $P$ is in days.\\
$^\dag$For Tethys and Dione, $a$ is in km, $P$ in days and $\delta\omega$ in deg/day.
\end{table}

Within standard gravitational theory, Kepler's law tells us that the left-hand side of Eq.~(\ref{Kepler2}) vanishes.
(More accurately, we may say that a statistical best-fit value is chosen for $A$ in such a way as to {\em force\/} the left-hand side as close to zero as possible for all systems observed.)
We turn this into a test of non-standard theory by summing the observational uncertainties associated with each of the quantities on the left-hand side to obtain an upper limit on the {\em right\/}-hand side of the equation.
For later convenience, we express this as
\begin{equation}
\left|\frac{\Delta_1}{m_1/m_2} + \Delta_2\right| \leq \epsilon_1 ,
\label{limit1}
\end{equation}
where, assuming uncorrelated errors,
\begin{equation}
\fl\epsilon_1 \equiv \left\{ \left[ \frac{\delta (m_\odot/m_1)}{m_\odot/m_1} \right]^2 + \left( 2\frac{\delta\omega}{\omega} \right)^2 + \left( 3\frac{\delta a}{a} \right)^2 + \left( 3\frac{\delta A}{A} \right)^2 + \left[ \frac{\delta(m_1/m_2)}{(m_1/m_2)^2} \right]^2 \right\}^{1/2} .
\label{epsilon1}
\end{equation}
(The first four terms in this equation are modified by a multiplicative factor $1+m_2/m_1$ but this has no effect on the results for any of the systems considered here.)
We apply Eq.~(\ref{epsilon1}) to eleven test-mass pairs as follows: \textit{Sun-planet} (with $m_1=m_{\odot}$ and $m_2=m_{\mbox{\scriptsize planet}}$; eight cases in all), \textit{Earth-Moon} (with $m_1=m_{\mbox{\scriptsize Earth}}$ and $m_2=m_{\mbox{\scriptsize Moon}}$), and \textit{Saturn-Trojan} (with $m_1=m_{\mbox{\scriptsize Saturn}}$ and $m_2=m_{\mbox{\scriptsize Tethys}}$ or $m_2=m_{\mbox{\scriptsize Dione}}$).
In general, most of the uncertainty comes from the semi-major axis ($\delta a$) term for the inner planets and Saturnian moons, while uncertainty in orbital frequency ($\delta\omega$) dominates for the outer planets.
Uncertainties in $A$ or $m_1/m_2$ are nearly always negligible  by comparison (uncertainty in the AU contributes 20\% of total uncertainty for the Earth, and the mass term figures at the 5\% level in the case of the Moon).

Results are summarized in Table~\ref{table-kepler}.
For the Earth we take $\delta a$ to be twice the uncertainty in $A$ following Ref.~\cite{Overduin2000}.
For Mars we take $\delta a$ to be twice the relevant range uncertainty of 300~m \cite{Folkner2011}.
For the other planets we use twice the maximum range uncertainty over the period 1950-2050, as plotted in Figs.~1-7 of Ref.~\cite{Folkner2011}.
(The small uncertainty for Saturn relative to Jupiter reflects Cassini's success vs. problems with the high-gain antenna during the earlier Galileo mission.)
For the Moon we take $\delta a$ to be twice the mean distance uncertainty, which is less than 60~cm from lunar laser ranging \cite{WilliamsDickey2003}.
For Tethys and Dione, data from Cassini give $\delta a$=20~m and 30~m respectively \cite{Antreasian2006}.
For mean motion uncertainty we take $\delta\omega$ from Ref.~\cite{WilliamsDickey2003} for the Moon, Ref.~\cite{HarperTaylor1993} for Tethys and Dione, and Ref.~\cite{Standish2004} for the planets.
All the figures for $\delta(m_1/m_2)$ come from Ref.~\cite{AstroAlmanac2012} except for those in the Saturn system, which are derived from Table~3 of Ref.~\cite{Jacobson2006}.

\section{Migration of Lagrange Points} \label{sec:lagrange}

\begin{figure}[t!]
\begin{center}
\includegraphics[width=80mm]{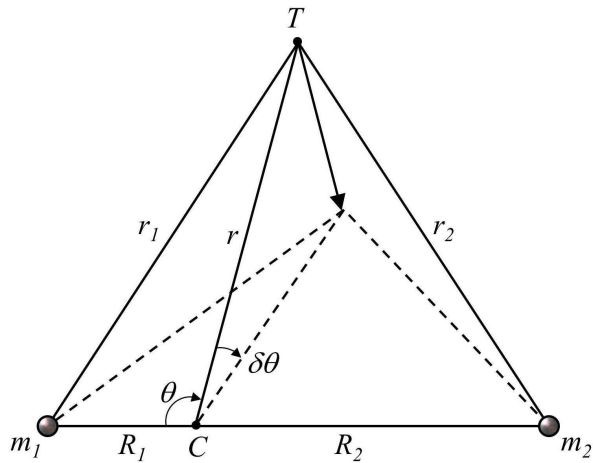}
\caption{Angular migration of the Lagrange point ($T$) through an angle $\delta\theta$ due to EP violations by $m_1,m_2$ and/or $m_T$.
$C$ is the center of rotation.
Both $R_1$ and the shift are exaggerated for emphasis; in practice $m_1 \gg m_2$ so that $R_1 \ll R_2$.}
\label{fig:lagrange}
\end{center}
\end{figure}

Kepler's law constrains one linear combination of $\Delta_1$ and $\Delta_2$, so we look to another observational quantity which depends on both parameters.
Lagrange points are stable or semi-stable points in the restricted three-body problem where a small test mass ($m_T$) will remain approximately motionless relative to the two larger masses ($m_1$ and $m_2$).
Two stable Lagrange points, called L4 and L5, exist $60^{\circ}$ in front of and behind each planet or moon ($m_2$) in its orbit around its parent body ($m_1$).
If $\Delta_1,\Delta_2$ and/or $\Delta_T$ are not zero, these points will be displaced in both the radial and angular directions, as originally shown by Nordtvedt \cite{Nordtvedt1968} and illustrated in Fig.~\ref{fig:lagrange}.
The angular shift offers better prospects as a probe of EP violation because of the  practical difficulty of obtaining accurate ranging data to distant asteroids.
In the case where $m_T\ll m_1$ and $m_T\ll m_2$ and all three masses violate the EP \cite{Overduin2000}:
\begin{equation}
\fl\delta\theta_L = \frac{R_1+R_2}{3\sqrt{3}(R_1^2+R_1R_2+R_2^2)} \left[(R_1+2R_2)(\Delta_1-\Delta_T) - (2R_1+R_2)(\Delta_2-\Delta_T)\right] .
\label{fullLagrange}
\end{equation}
Since $m_1\gg m_2$ in all the cases we consider here, we may take $R_1\ll R_2$.
If in addition $m_1$ and $m_2$ are similar bodies, with $m_T$ different in composition such that $\Delta_T\ll\Delta_1$ and $\Delta_T\ll\Delta_2$, then
\begin{equation}
\delta\theta_L = \frac{1}{3\sqrt{3}}(2\Delta_1 - \Delta_2) .
\label{simplifiedLagrange}
\end{equation}
Such a situation could apply in the case of the Trojan asteroids in the Sun-Jupiter system, for example, or the recently discovered Trojan companions of Neptune.

Other possibilities could be explored as well.
For example, if $m_2$ and $m_T$ were compositionally similar, but both different from $m_1$, then one might look for a constraint on the difference $\Delta_1-\Delta_T$.
Such a situation might be used to model the Trojan satellites of Saturn, Mars and the Earth.
For the present we follow Refs.~\cite{Nordtvedt1968,Orellana1993,Overduin2000} in adopting Eq.~(\ref{simplifiedLagrange}) for our analysis of Lagrange point constraints.
This gives us a linearly independent constraint on many of the same pairs of test bodies already considered in Section~\ref{sec:kepler}.
Following the same approach, we use upper limits on observational uncertainty in $\theta_L$ to set an upper limit on $\Delta_1$ and $\Delta_2$ so that
\begin{equation}
\left|\Delta_1 - {\textstyle \frac{1}{2}}\Delta_2\right| \leq \epsilon_2 ,
\label{limit2}
\end{equation}
where
\begin{equation}
\epsilon_2 \equiv {\textstyle \frac{3\sqrt{3}}{2}} \delta\theta_L .
\label{epsilon2}
\end{equation}
Here $\delta\theta_L$ is an estimate of uncertainty in the angular position of the Lagrange points.
The locations of L4 and L5 must, of course, be inferred in practice from observations of the objects that accumulate there over time.
Nearly 6000 Trojan asteroids have been detected around Jupiter \cite{mpc}, out of a total population estimated at more than 300\,000 \cite{Jewitt2000}.
Nine Trojans have been discovered near Neptune's Lagrange points \cite{Sheppard2006,Sheppard2010,Guan2012,Parker2013}, where the total population is thought to be even larger.
Mars has three known Trojan companions \cite{Connors2005}, and the Earth one \cite{Connors2011}.
Finally, while simulations suggest that Saturn and Uranus do not harbor large numbers of stable Trojans \cite{Alexandersen2013}, two of Saturn's {\em moons\/} do have smaller Trojan companions: Telesto and Calypso in the orbit of Tethys, and Helene and Polydeuces in the orbit of Dione \cite{Murray2005}.

To locate the mean angular position of these objects with sufficient precision for EP tests can pose a significant challenge.
Older observations are subject to larger random scatter than more recent ones.
There are several potential sources of systematic error, including observational selection effects and the nonuniform distribution of the Trojans, which may not necessarily cancel themselves out over time.
But the greatest source of uncertainty for most of the systems we consider is {\em libration\/}.
Trojans do not simply congregate near L4 and L5; rather they wander around these points with libration periods $\Tlib$ that can greatly exceed the timescale $\Tobs$ over which the Trojans themselves have been observed.
The task of locating the center of libration for such objects is akin to determining the phase of a sine wave from an arc of observations covering only a fraction of the wavelength.
The error in such a procedure goes as approximately $t^{-2}$ for short observation times $t$ relative to the period.
When $\Tobs\ll\Tlib$ we therefore take
\begin{equation}
\delta\theta_L = \frac{1}{\sqrt{n}}\left(\frac{\Tlbar}{\Tobar}\right)^2\delta\bar{\theta}_T ,
\label{trojanShort}
\end{equation}
where $\Tlbar$ and $\Tobar$ are the mean libration period and observation time for $n$ Trojans whose mean angular orbit uncertainty is $\delta\bar{\theta}_T$.
Current and regularly updated values for $\delta\theta_T$ are now available online for most objects; e.g., on the AstDyS-2 website for asteroids \cite{astdys2}.
We take $\delta\theta_T$ to be the rms value of each object's ephemeris uncertainty ellipse.
If observations are available for more than one Trojan, we choose the value of $n$ so as to minimize $\delta\theta_L$.
This may mean using only a small fraction of the known population.
Trojans which have been observed for insufficiently long relative to their libration periods are discarded since the net increase in $\Tlbar/\Tobar$ more than outweighs the root-$n$ reduction in uncertainty.

\begin{figure}[t!]
\begin{center}
\includegraphics[width=120mm]{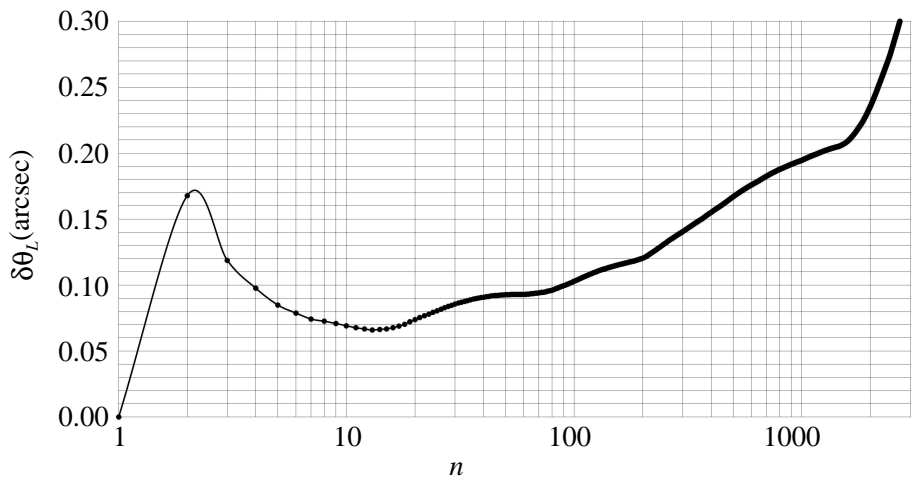}
\caption{Angular position uncertainty $\delta\theta_L$ of Jupiter's Lagrange points, as given by Eq.~(\ref{trojanShort}), evaluated as a function of the number $n$ of Trojan asteroids used (ordered by time since discovery and assuming for simplicity that $\delta\bar{\theta}_T\approx$~const beyond $n>12$; see text for discussion).}
\label{fig:howManyTrojans}
\end{center}
\end{figure}

Jupiter presents a particularly interesting case.
The current average $1\sigma$ rms orbit uncertainty for the twelve oldest Jovian Trojans is 0.08~arcsec \cite{astdys2}.
Their average libration period is $\Tlbar=154$~yr \cite{BienSchubart1987} and they have been observed for an average of $\Tobar=91$~yr \cite{mpc}.
Eq.~(\ref{trojanShort}) then gives $\delta\theta_L < 0.07$~arcsec.
(By comparison, the typical observing resolution for {\it individual\/} observations of Jovian Trojans varies between 0.3-0.5~arcsec \cite{Fernandez2003} and 0.8-1.0~arcsec \cite{Jewitt2000}.)
This in Eq.~({\ref{epsilon2}}) leads to $\epsilon_2<8\times10^{-7}$ for the Sun-Jupiter system.
No benefit is derived by incorporating additional Jovian Trojans, as demonstrated in Fig.~\ref{fig:howManyTrojans} where the angular position uncertainty $\delta\theta_L$ is plotted as a function of the number $n$ of asteroids considered, ordered by the time elapsed since discovery (longest to shortest).
The new additions merely add to the statistical noise because they have not been observed for long enough relative to their libration periods.
We have assumed that mean orbit uncertainty $\delta\bar{\theta}_T$ remains constant beyond $n>12$, which is conservative insofar as newly discovered asteroids will tend to have larger orbit uncertainties than those with long observation histories.
Thus the actual curve probably climbs more steeply with $n$ than shown here.

Similar considerations apply to the Trojan companions of Neptune, Mars and the Earth.
The mean orbit uncertainty for the nine Neptune Trojans observed to date is $\delta\bar{\theta}_T=$20~arcsec \cite{astdys2}.
They have been observed for an average of 7~yrs, far less than their mean libration period is 9400~yrs \cite{Almeida2009,Lykawka2011,Guan2012,Parker2013}.
The resulting formal limit of $\epsilon_2<200$ in Table~\ref{table-lagrange} indicates that {\it no\/} useful information can be gleaned from these objects as to the actual location of Neptune's Lagrange points.
\begin{table}[t!]
\caption{Limits from migration of the Lagrange points.}
\label{table-lagrange}
\begin{indented}
\item[]\begin{tabular}{@{}llllll}
\br
Pair ($m_{\mbox{\tiny 1}}-m_{\mbox{\tiny 2}}$) & $n$ & $\bar{T}_{\mbox{\tiny lib}}$ (yrs) & $\bar{T}_{\mbox{\tiny obs}}$ (yrs) & $\delta\bar{\theta}_T$ (arcsec) & $\epsilon_{\mbox {\tiny 2}}$\\
\mr
Sun-Earth & 1 & 400 & 2.5 & 2 & $8\times10^{-1}$ \\
Sun-Mars & 3 & 1400 & 17 & 0.05 & $2\times10^{-3}$ \\
Sun-Jupiter & 12 & 150 & 92 & 0.08 & $8\times10^{-7}$ \\
Sun-Neptune & 9 & 9400 & 7 & 20 & $2\times10^2$ \\
Saturn-Tethys & 2 & 1.9 & 33 & 20 & $2\times10^{-4}$ \\
Saturn-Dione & 2 & 2.1 & 21 & 10 & $9\times10^{-5}$ \\
\br
\end{tabular}
\end{indented}
\end{table}
(This is reasonable, given that they have been discovered as a result of intentional searches in the regions around L4 and L5.
The Trojan designation is conferred after numerical simulations of objects with similar orbital characteristics remain co-orbital with their parent body over a significant fraction of the age of the solar system.)
The Martian Trojans have a mean orbit uncertainty $\delta\bar{\theta}_T=$0.05~arcsec \cite{astdys2}.
But again they have been observed for an average of only 17 yr \cite{mpc}, versus a mean libration period of 1400 yr \cite{Connors2005,Guan2012}.
The resulting limit on EP violation of order $\epsilon_2<2\times 10^{-3}$ is of marginal interest.
For the Earth, numbers are comparable.
Newly discovered companion 2010~TK$_7$ has a current orbit uncertainty of $\sim 2$~arcsec \cite{astdys2} but librates with a period of 400 yr \cite{Connors2011}, resulting in a formal limit of $\epsilon_2<0.8$.

A different situation prevails when a Trojan satellite has been observed for significantly longer than its libration period, as with Saturn's Trojan moons.
For these cases the center of libration can be established with more confidence, and we take
\begin{equation}
\delta\theta_L = \delta\bar{\theta}_T/\sqrt{n} .
\label{trojanLong}
\end{equation}
Cassini observations currently imply rms orbit uncertainties of less than about 30~km for Calypso and Telesto and 20~km for Helene and Polydeuces \cite{JacobsonEmail}.
Since Calypso and Telesto orbit Saturn (together with Tethys) at $a=294\,000$~km, while Helene and Polydeuces share Dione's orbit at $a=377\,000$~km, these numbers translate into angular uncertainties of $\delta\bar{\theta}_T=20$~arcsec (Tethys) and 10~arcsec (Dione), as indicated in Table~\ref{table-lagrange}.
The libration periods for all four moons are approximately two years, while Telesto, Calypso and Helene have all been observed for over 30~years, and Polydeuces for nearly 10 \cite{Christou2007,Spitale2006}.
These numbers in Eq.~(\ref{trojanLong}) produce upper limits of $\epsilon_2<2\times 10^{-4}$ and $1\times10^{-4}$ on the Saturn-Tethys and Saturn-Dione systems respectively.

\section{Orbital Polarization} \label{sec:nordtvedt}

\begin{figure}[t!]
\begin{center}
\includegraphics[width=80mm]{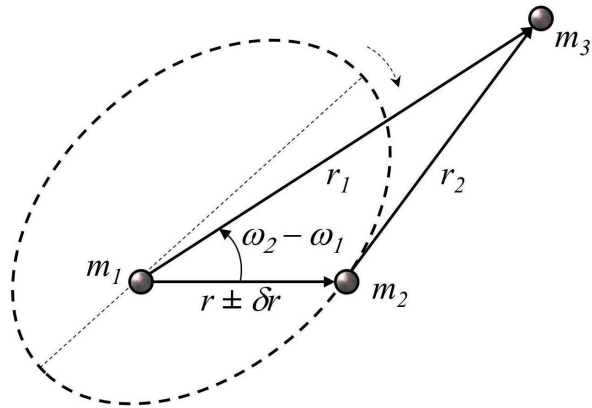}
\caption{Orbital polarization (Nordtvedt effect). Bodies $m_1$ and $m_2$ fall with different accelerations toward $m_3$, producing periodic oscillations in distance $r$ at the synodic frequency $\omega_2-\omega_1$.}
\label{fig:nordtvedt}
\end{center}
\end{figure}

The modified Kepler's law~(\ref{Kepler1}) depends on the sum of $\Delta$ parameters; while migration of the Lagrange points, Eq.~(\ref{simplifiedLagrange}), depends on the difference.
In principle the combination can give us limits on the individual $\Delta$ parameters.
However, Trojan-based constraints are weak in many cases.
For most pairs of bodies a stronger complementary limit on EP violation can be obtained using orbital polarization (also known as the Nordtvedt effect \cite{Nordtvedt1968,Nordtvedt1968b,Nordtvedt1968c,Nordtvedt1970}), whereby two masses ($m_1$ and $m_2$) with different values of $\Delta$ fall toward a third ($m_3$) with different accelerations (Fig.~\ref{fig:nordtvedt}).

The distance between $m_1$ and $m_2$ then undergoes periodic oscillations at the synodic frequency $\omega_2-\omega_1$, where $\omega_1$ is the orbital frequency of $m_1$ about $m_3$ (or vice versa), and $\omega_2$ is the orbital frequency of $m_2$ about $m_1$.
This has the effect of aligning or ``polarizing'' the orbit of $m_2$ about $m_1$ along the direction either toward $m_3$ (if $\Delta_1>\Delta_2$) or away from $m_3$ (if $\Delta_1<\Delta_2$).
The maximum amplitude of the oscillations is given by \cite{Overduin2000,Will1993}
\begin{equation}
\delta r = (\Delta_1-\Delta_2)A_{\mbox{\tiny EP}} ,
\end{equation}
where 
\begin{equation}
A_{\mbox{\tiny EP}} = \left[ 
   \frac{1+2\omega_2/(\omega_2-\omega_1)}{2 \, (\omega_2/\omega_1)-1} 
   \right] r_1 ,
\label{AEP}
\end{equation}
and $r_1$ is the mean distance between $m_1$ and $m_3$.
In some cases (e.g., the Earth-Moon system) the oscillations are magnified by tidal effects.

Following the same approach as in the preceding sections, we isolate the effect of these oscillations on the difference in $\Delta$ terms as
\begin{equation}
|\Delta_1-\Delta_2| \leq \epsilon_2 ,
\label{limit3}
\end{equation}
where
\begin{equation}
\epsilon_2 \equiv \frac{\delta r}{r_1} \left[ 
   \frac{(P_1-P_2)(2P_1-P_2)}{P_2(3P_1-P_2)} 
   \right] ,
\label{limit3b}
\end{equation}
and where we have re-expressed the standard result in terms of orbital periods rather than frequencies for convenience.
The observational uncertainty $\delta r$, together with values of $r_1,P_1$ and $P_2$, then impose experimental upper limits on $|\Delta_1-\Delta_2|$ for various systems.
Strong limits are obtainable in principle if $\delta r$ is small in comparison to the ``lever arm'' $r_1$---and also in cases where $P_2\approx P_1$ or $P_2\approx 2P_1$.

We apply this method to situations of three kinds.
First, the Earth ($m_1$) and Moon ($m_2$) falling toward the Sun ($m_3$).
This was the original, and remains the definitive application of orbital polarization, thanks to the precision with which the Earth-Moon is known from lunar laser ranging.
For the second and broader class of systems, we use as a baseline the distance between the Sun ($m_1$) and a ``primary'' planet ($m_2$), both undergoing mutual accelerations toward a second or ``perturbing'' planet ($m_3$).
(The period of $m_3$ and $m_1$ about their common barycenter are of course the same.)
This was originally applied by Nordtvedt to the case where Jupiter acts as a perturber on the Sun-Earth system; one motivation being that Jupiter might be likeliest to violate the EP by virtue of its significant gravitational self-energy \cite{Nordtvedt1970}.
%(The resulting constraints were compared with those obtained from the migration of the Jovian Lagrange points in Ref.~\cite{Overduin2000}.)
Theorists now consider many other mechanisms for EP violation, and ranging distances to most of the planets have gained tremendously in precision thanks to missions such as Cassini.
We therefore apply the same method systematically here to {\em all\/} planetary combinations in the solar system.
That is, for each primary ($m_2$) we treat every other planet ($m_3$) as a possible perturber, and choose for our limit the one that best constrains the Sun-primary pair.
Finally, as a third application we consider cases in which Saturn ($m_1$) and its Trojan moons Tethys and Dione ($m_2$) fall with possibly different accelerations toward the Sun ($m_3$).
This is motivated theoretically by the very different compositions of the two moons, and observationally by the availability of high-precision ranging data from the Cassini mission.

Our results are summarized in Table~\ref{table-nordtvedt}.
\begin{table}[t!]
\footnotesize
\caption{Limits from orbital polarization.}
\label{table-nordtvedt}
\begin{indented}
\item[]\begin{tabular}{@{}llllll}
\br
$m_{\mbox{\tiny 1}}$ & $m_{\mbox{\tiny 2}}$ & Best $m_{\mbox{\tiny 3}}$ & $\delta r$ (km) & $\epsilon_{\mbox {\tiny 2}}$ & Synodic period (yrs) \\
\mr
Sun & Mercury & Venus & 5 & $4\times10^{-8}$ & 0.396 \\
Sun & Venus & Earth & 1 & $2\times10^{-9}$ & 1.60 \\
Sun & Earth & Venus & 0.015 & $2\times10^{-11}$ & 1.60 \\
Sun & Mars & Earth & 1.5 & $5\times10^{-10}$ & 2.14 \\
Sun & Jupiter & Saturn & 50 & $3\times10^{-8}$ & 19.9 \\
Sun & Saturn & Uranus & 1.5 & $6\times10^{-10}$ & 45.3 \\
Sun & Uranus & Neptune & 1000 & $1\times10^{-7}$ & 172.7 \\
Sun & Neptune & Uranus & 5000 & $4\times10^{-8}$ & 172.7 \\
Earth & Moon & Sun & 6.7$^{\ast}$ & $2\times10^{-13}$ & 29.53$^{\dag}$ \\
Saturn & Tethys & Sun & 0.10 & $3\times10^{-7}$ & 1.888$^{\dag}$ \\
Saturn & Dione & Sun & 0.15 & $3\times10^{-7}$ & 2.738$^{\dag}$ \\
\br
\end{tabular}
\\$^\ast$For the Moon, $\delta r$ is in mm.
\\$^\dag$For the Moon, Tethys and Dione, synodic period is in days.
\end{indented}
\end{table}
We discuss a few illustrative cases here.
As expected, the strongest constraints arise in the Earth-Moon-Sun case.
Using recent figures $A_{\mbox{\tiny EP}}=2.992\times10^{13}$~mm (including tidal effects) and $\delta r\leqslant 6.70$~mm from lunar laser ranging \cite{Williams2012}, we obtain an upper bound $\epsilon_2=2\times10^{-13}$ that is twice as strong as that in Ref.~\cite{Overduin2000}.

For the planets, the strongest limits in each case arise when $P_2$ is closest to $P_1$; that is, when the perturbing planet is in an orbit adjacent to the primary.
This is as expected on the basis of Eq.~(\ref{limit3b}).
The numerical strength of each best-case limit is then determined primarily by the uncertainty $\delta r$ in the distance between the Sun ($m_1$) and primary body ($m_2$).
As a conservative estimate for this quantity we follow Ref.~\cite{Overduin2000} in adopting a value of five times the maximum ephemeris range uncertainty to each planet, as listed in Table~\ref{table-kepler}.
(For the case of the Earth itself, we use five times the uncertainty in the AU; for all other cases the uncertainty in Sun-Earth distance contributes negligibly to $\delta r$.)
There are large {\em classical\/} perturbations in each planet's distance from the Sun at the same synodic frequency as the putative EP signal, but these can be accurately compensated for since the mass ratios $m_1/m_2$ are known to sufficient precision in every case.

The strongest planetary limit of $\epsilon_2=2\times10^{-11}$ is found for the Earth, using Venus as a perturber.
(Using Mars instead weakens this slightly to $3\times10^{-11}$.
Using Jupiter as the perturber, as was done in Refs.~\cite{Overduin2000,Nordtvedt1970}, leads to a considerably weaker constraint of $1\times10^{-10}$ for the Earth-Sun pair.)
Also noteworthy are the limits of order $\sim10^{-9}$ obtained for Mars and Jupiter using the Earth and Saturn as perturbers, respectively.
(The strong limit on Saturn reflects excellent Cassini ranging data to that planet.)
The upper limit of $\epsilon_2=3\times10^{-8}$ for Jupiter (using Saturn as a perturber) is weaker, but still thirty times stronger than the comparable constraint on the Sun-Jupiter combination from migration of the Lagrange points in Table~\ref{table-lagrange}.

For the Saturnian moons, finally, we use the Sun as a perturber so that $r_1=9.5$~AU and take $\delta r = 5\delta a$ as usual (where Cassini limits on $\delta a$ are listed in Table~\ref{table-kepler}).
This leads to upper limits $\epsilon_2=3\times10^{-7}$ in both cases.
For both moons there is a gain in sensitivity due to the large ``lever arm'' $r_1\gg\delta r$.
However, this geometrical factor is more than offset by the fact that $\epsilon_2$ in Eq.~(\ref{limit3}) is roughly proportional to a factor of $P_1/P_2$ where $P_1$ (the period of Saturn around the Sun) in this case is much greater than $P_2$ (the orbital period of either moon around Saturn).
The resulting upper limits on $\epsilon_2$ are weaker than those of most of the planets, but still two to three orders of magnitude stronger than the comparable constraints on the Saturnian moons from migration of the Lagrange points in Table~\ref{table-lagrange}.
At the same time, the level of agreement between these two completely independent ways of constraining EP violation for both the Saturnian and Jovian cases serves as a useful consistency check.

\section{Limits on Individual Bodies} \label{sec:individual}

Our limits on EP violation by pairs of solar-system bodies to this point are given by Eq.~(\ref{limit1}) from the modified Kepler's law, Eq.~(\ref{limit2}) from the migration of stable Lagrange points and Eq.~(\ref{limit3}) from orbital polarization (Nordtvedt effect).
These can be summarized in the form of two inequalities:
\begin{equation}
| \frac{\Delta_1}{c_1}+\Delta_2 | < \epsilon_1 \;\;\; , \;\;\;
| \Delta_1-c_2\Delta_2 | < \epsilon_2 ,
\label{bothLimits}
\end{equation}
where
\begin{eqnarray}
c_1 & \equiv & \frac{m_1}{m_2} \;\; \mbox{ (Kepler) } \nonumber \\
c_2 & \equiv & \left\{ \begin{array}{ll}
   1/2 & \mbox{ (Lagrange) } \\
   1   & \mbox{ (Nordtvedt) }
   \end{array} \right. \nonumber
\end{eqnarray}
Eqs.~(\ref{bothLimits}) can be squared and combined to extract algebraic expressions for upper limits on $\Delta_1$ and $\Delta_2$ \cite{Overduin2000}.
Alternatively, one can invert and consider all four possible cases separately.
In each case one finds that
\begin{equation}
|\Delta_1| < |c_2\,\epsilon_1 + \epsilon_2| \;\;\; , \;\;\;
|\Delta_2| < |\epsilon_1 + \epsilon_2/c_1| .
\label{individualLimits}
\end{equation}
This result neglects a multiplicative factor of $(1+c_2/c_1)^{-1}$ on some terms, an approximation that overestimates the uncertainty by less than 1\% for all the systems considered here (reaching a maximum of 1/81 for the Earth-Moon case).
We then substitute the relevant values of $\epsilon_1$ and $\epsilon_2$ from Tables~\ref{table-kepler}, \ref{table-lagrange} and \ref{table-nordtvedt} into Eqs.~(\ref{individualLimits}) and select the best limit for each solar-system body.
\begin{table}[t!]
\footnotesize
\caption{Limits for individual bodies.}
\label{table-individual}
\begin{indented}
\item[]\begin{tabular}{@{}lll}
\br
Body & $\Delta_{\mbox{\scriptsize max}}$ & Source$^{\ast}$ \\
\mr
Sun & $2\times10^{-10}$ & Sun-Earth (K+N) \\
Mercury & $1\times10^{-7}$ & Sun-Mercury (K+N) \\
Venus & $1\times10^{-8}$ & Sun-Mercury (K+N) \\
Earth & $1\times10^{-10}$ & Sun-Earth (K+N) \\
$\hspace{5mm}$ Moon & $9\times10^{-9}$ & Earth-Moon (K+N) \\
Mars & $8\times10^{-9}$ & Sun-Mars (K+N) \\
Jupiter & $9\times10^{-8}$ & Sun-Jupiter (K+L/N) \\
Saturn & $9\times10^{-8}$ & Sun-Saturn (K+N) \\
$\hspace{5mm}$ Tethys & $2\times10^{-7}$ & Saturn-Tethys (K+L/N) \\
$\hspace{5mm}$ Dione & $2\times10^{-7}$ & Saturn-Dione (K+L/N) \\
Uranus & $5\times10^{-7}$ & Sun-Uranus (K+N) \\
Neptune & $2\times10^{-6}$ & Sun-Neptune (K+N) \\
\br
\end{tabular}
\\$^\ast$(K=Kepler, L=Lagrange and N=Nordtvedt)
\end{indented}
\end{table}
Results are listed in Table~\ref{table-individual}, where ``K,'' ``L'' and ``N'' refer to limits obtained from the modified Kepler's law, migration of stable Lagrange points and orbital polarization (Nordtvedt effect) respectively.
The strongest limits in every case come from combining Kepler's third law with orbital polarization.
However, the combination of Kepler plus Lagrange comes close in the case of Mars, and gives {\em equally\/} strong results in the cases of Jupiter and the Saturnian moons.
This may be understood from Eqs.~(\ref{individualLimits}), where it is seen that $\Delta_2$ is essentially equivalent to $\epsilon_1$ from Kepler's third law.
The contribution of the Lagrange limit $\epsilon_2$ is suppressed by a factor of $1/c_1$.

Some solar-system bodies are constrained in more than one way.
Upper limits for the Sun, for example, come from every Sun-planet pair.
We select the strongest constraint in each case.
For the Sun this comes from the Sun-Earth combination, $\Delta_{\mbox{\footnotesize Sun}}<2\times10^{-10}$.
Similarly, limits for the Earth come from both the Sun-Earth pair (where the Earth plays the role of $m_2$) and the Earth-Moon pair (where it is $m_1$).
The former gives the stronger limit in this case, $\Delta_{\mbox{\footnotesize Earth}}<1\times10^{-10}$.
(For comparison the Earth-Moon combination gives $\Delta_{\mbox{\footnotesize Earth}}<1\times10^{-8}$.)
Similar comments apply to Saturn, which is constrained by both the Sun-Saturn and Saturn-Tethys/Dione combinations (the former giving $\Delta_{\mbox{\footnotesize Sat}}<9\times10^{-8}$ while the latter both imply $\Delta_{\mbox{\footnotesize Sat}}<2\times10^{-7}$).
These results for the Earth and Sun are both fifty times stronger than those previously reported in Ref.~\cite{Overduin2000}.

Numerically, the other constraints in Table~\ref{table-individual} range from order  $10^{-8}$ (for the Moon, Mars and Venus) to $10^{-7}$ (Mercury, Jupiter, Saturn, Tethys and Dione) and finally $10^{-6}$ (Uranus, Neptune).
In the case of the Moon this result is comparable to that previously reported in Ref.~\cite{Overduin2000}, while the limit for Jupiter is an order of magnitude stronger.
For all the other bodies, these are the first such limits to be reported.

The sole $\Delta_1$-based limit here, that for the Sun, reflects uncertainties in both the Kepler ($\epsilon_1$) and Nordtvedt ($\epsilon_2$) methods in roughly equal proportion, as shown by Eqs.~(\ref{individualLimits}).
By contrast, the best limits for all the other bodies are $\Delta_2$-based, meaning that they are largely determined by the observational uncertainties in Kepler's third law alone.
This is important, as it points to the most effective way to strengthen similar solar-system-based EP tests in the future.
For the present, we would regard the results in Table~\ref{table-individual} as valid to order of magnitude (see Discussion below).

\section{Application and Elemental Limits} \label{sec:elemental}

The constraints derived above apply to any theory in which different bodies may violate the EP in essentially independent ways.
One example occurs in Kaluza-Klein gravity, where the gravitational field around a static, spherically-symmetric central mass is commonly modeled with a generalization of the Schwarzschild metric of general relativity known as the soliton  metric \cite{OW97}.
In this theory it may be shown that $\Delta\approx-b/2$ where $b$ is a free parameter of the soliton metric, related to the curvature of the extra dimension in the vicinity of the central mass (standard general relativity is recovered on four-dimensional hypersurfaces as $b\rightarrow0$).
This has led to the strongest current constraints on Kaluza-Klein gravity with the soliton metric, $|b|<2\times10^{-8}$ for the Earth, Sun and Moon and $|b|<2\times10^{-6}$ for Jupiter \cite{Overduin2000}.
These bounds are marginally consistent with theory, since theoretical calculations \cite{Liu2000} suggest that $b$ might range from $\sim10^{-8}-10^{-2}$ in gravitationally condensed objects like planets, but take values as large as $\sim0.1$ in more diffuse matter distributions such as galaxy clusters.

The new results in Table~\ref{table-individual} further strengthen these bounds and extend them to more solar-system bodies.
The previously reported upper bound on $b$ for Jupiter goes down by one order of magnitude, while those for the Earth and Sun drop by two.
The new limit for the Earth, $|b|<2\times10^{-10}$, though less direct, is orders of magnitude stronger than that recently imposed by measurements of gyroscope precession in low-earth orbit \cite{OEW13} and casts particular doubt on the applicability of the soliton solution within higher-dimensional relativity.
The other solar-system bodies we have considered here are all constrained for the first time by these results.
Any other theories that predict explicit EP violation on macroscopic scales would be subject to similar constraints.

Most theories that involve EP violation in principle do not make such concrete predictions for $\Delta$.
They do, however, agree that the degree of EP violation will depend in some way on {\em composition} \cite{Damour1994,Damour2002,Carroll1998,Chen2005,Antoniadis2001,Dvali2002,Wetterich2003,Khoury2004,Mota2007,Capozziello2008,Kostelecky2009,Kostelecky2010}.
This is because of the presence of new fields that couple non-universally but with gravitational strength to the constituents of the standard model.
In the absence of a definitive theory, the standard way to characterize such EP violations is to define a phase space of the most plausible observables, such as baryon number, neutron excess and electrostatic binding energy.
%(Fig.~\ref{fig:testMaterials}).
%\begin{figure}[t!]
%\begin{center}
%\includegraphics[width=120mm]{fig4color.eps}
%\caption{Various elements and compounds plotted in three plausible dimensons of ``EP-violating phase space'' defined by baryon number $N+Z$, neutron excess $N-Z$ and electrostatic binding energy $[\propto Z(Z-1)]$, all normalized by atomic mass $A$.
%The most common isotopes are indicated with circles; lighter isotopes are squares; heavier isotopes are triangles and diamonds; compounds are patterned (after Ref.~\cite{Overduin2012}).}
%\label{fig:testMaterials}
%\end{center}
%\end{figure}
In designing a experiment, one hopes to drop test materials that span the largest possible volume in this space, while also ensuring that any signal seen is as robust as possible \cite{Blaser2003,Overduin2012}.
Here we extend this phenomenological approach to the solar system, combining our upper limits on EP violation by the Sun, planets and satellites with compositional data on each body to extract upper limits on $\Delta$ for individual constituent elements themselves.

Given the theoretical uncertainties, we take the simplest possible approach in which EP violation by a macroscopic body is due entirely to a single constituent, neglecting possible internal cancelation or other effects.
This means that we are effectively comparing the acceleration of that one element to the rest of the periodic table.
Thus, for example, the bulk composition of the Sun consists of 72\% hydrogen and 27\% helium by mass \cite{Grevesse2010}.
Assuming that any EP violation by the Sun can effectively be associated with a single element, our limit of $\Delta_{\mbox{\footnotesize Sun}}<2\times10^{-10}$ for the Sun would imply that $\Delta_{\mbox{\footnotesize H}}<2\times10^{-10}$ for hydrogen or $\Delta_{\mbox{\footnotesize He}}<6\times10^{-10}$ for helium.
These are in fact our best bounds on these two elements from solar-system observations.
Jupiter's atmosphere consists of 76\% hydrogen and 24\% helium by mass, and the comparable fractions for Saturn are 79\% and 21\% \cite{Atreya2003}, but our upper limits on $\Delta$ for these planets are much weaker than that for the Sun.
Similarly for Uranus and Neptune, estimated to consist of 10\% atmospheric hydrogen and helium, plus a core of 25\% silicate rock and 65\% water ice \cite{Guillot1999}.

Limits derived in this way for hydrogen, helium and the four major constituent elements of the terrestrial planets (oxygen, magnesium, silicon and iron) and icy satellites are listed in Table~\ref{table-elements}.
\begin{table}[t!]
\caption{Derived limits for selected elements}
\label{table-elements}
\begin{indented}
\item[]\begin{tabular}{@{}rll}
\br
Element & $\Delta_{\mbox{\scriptsize max}}$ & Source body \\
\mr
H & $2\times10^{-10}$ & Sun \\
He & $6\times10^{-10}$ & Sun \\
O & $5\times10^{-10}$ & Earth \\
Mg & $9\times10^{-10}$ & Earth \\
Si & $9\times10^{-10}$ & Earth \\
Fe & $4\times10^{-10}$ & Earth \\
\br
\end{tabular}
\end{indented}
\end{table}
For the Earth we adopt mass fractions of 32\%~Fe, 30\%~O, 16\%~Si and 15\%~Mg \cite{McDonough1995}.
Comparable numbers for the Moon are 8\%~Fe, 44\%~O, 22\%~Si and 21\%~Mg
\cite{Warren2005}.
For Mars we use 27\%~Fe, 34\%~O, 17\%~Si and 14\%~Mg \cite{Lodders1996}.
Corresponding figures for Venus are 30\%~Fe, 34\%~O, 15\%~Si and 15\%~Mg
while Mercury has 63\%~Fe, 14\%~O, 7\%~Si and 7\%~Mg \cite{Taylor1982}.

Saturn's icy satellites Tethys and Dione constitute a particularly tempting EP test case, %(Fig.~\ref{fig:saturnMoons}),
%\begin{figure}[t!]
%\begin{center}
%\includegraphics[width=\columnwidth]{fig5color.eps}
%\caption{Saturn with satellites Dione (left inset) and Tethys (center) as imaged by Voyager~1 in 1980 and Voyager~2 in 1981 respectively.
%Cassini has confirmed that Tethys is almost pure water ice while Dione is approximately one-half silicate rock by mass \cite{Matson2009}, making these two moons with their Trojan companions Calypso, Telesto, Helene and Polydeuces a particularly interesting EP test case.}
%\label{fig:saturnMoons}
%\end{center}
%\end{figure} 
since Cassini has confirmed that one (Tethys) consists of 93\% water ice by mass while the other (Dione) is 50\% silicate rock \cite{Matson2009}.
(In elemental terms these numbers translate into 86\%~O and 4\%~Si by mass for Tethys vs. 74\%~O and 20\%~Si for Dione.)
As both have Trojan companions, they can be constrained not only by the combination of Kepler's third law and orbital polarization, but by the migration of their stable Lagrange points as well.
Moreover, we have excellent data on both moons and their Trojan companions from Cassini.
However, for the reasons discussed in Sec.~\ref{sec:individual}, the upper limits on $\Delta$ for both bodies are still comparatively weak.
The best elemental limits in every case turn out to be those derived from the Earth, whose upper bound on $\Delta$ is two or more orders of magnitude stronger than any of the other terrestrial planets.

\section{Summary and Discussion} \label{sec:discussion}

We have looked for the constraints imposed by solar system data on theories in which the ratio of gravitational to inertial mass differs from unity by a factor $\Delta$ which may in principle differ from body to body.
For two objects characterized by $\Delta_1$ and $\Delta_2$, upper bounds on the sum $|\Delta_1/c_1+\Delta_2|$ are set by Kepler's third law, while the difference $|\Delta_1-c_2\Delta_2|$ is constrained by the position of Lagrange libration points and orbital polarization in the field of a third body (the Nordtvedt effect).
(Here $c_1$ and $c_2$ are known constants.)
Combining these results, we have extracted independent upper limits on $\Delta$ for the Sun, Moon, planets and Saturnian moons Tethys and Dione using experimental data on their mean motions and positions as well as those of their Trojan companions where applicable.
We find that $\Delta\lesssim10^{-10}$ for the Earth and Sun, $\Delta\lesssim10^{-8}$ for the Moon, Mars and Venus, $\Delta\lesssim^{-7}$ for Mercury, Jupiter, Saturn, Tethys and Dione, and  $\Delta\lesssim10^{-6}$ for Uranus and Neptune.

As a test case, we have applied our results to Kaluza-Klein gravity, in which $\Delta$ depends on a metric parameter related to the curvature of an extra dimension near the central mass.
Our upper bounds on this parameter are orders of magnitude stronger than existing limits from any other tests, and confirm earlier conclusions that a fifth dimension, if any, plays no significant dynamical role in the solar system.

We have combined our limits with data on the composition of each solar-system body to obtain constraints on EP violation by individual constituent elements, under the assumption that a single element dominates in each case.
The resulting upper limits on $\Delta$ for hydrogen, helium, iron, oxygen, silicon and magnesium are of order $10^{-9}-10^{-10}$.

There is an important statistical caveat to these results.
As uncertainties in our orbital and other parameters, we have implicitly relied on residuals from published fits to a fixed number of ephemeris solution parameters.
These fits do not generally incorporate a different degree of EP violation for each solar-system body.
(They are typically sensitive to at most a single EP-violating parameter $\eta$.)
We have, in other words, relied on more degrees of freedom than are actually present in the solutions.
This is not necessarily a problem, but will tend to underestimate our uncertainties.
The results least affected will be those based on the lunar Nordtvedt effect, for which at least one EP-violating term is explicitly included in the solution sets.
Our other results may be less robust in comparison.
It would be of interest to incorporate additional independent parameters for EP violation into the standard ephemeris models.

Given the uncertainties, our results should be seen as illustrative rather than definitive.
They are three to seven orders of magnitude weaker than the best existing constraints on EP violation from torsion balances or lunar laser ranging.
Space-based free-fall experiments should produce even stronger bounds.
However, in the context of modern unified theories it may be at least as important to explore a broad range of test materials as to so with the greatest possible sensitivity.
In this respect, solar-system tests offer a diversity of composition unobtainable in any other way.
Moreover, they do so at comparatively little cost.

\ack
We thank B.~Eney and A.~Storrs for discussions along with R.J.~Adler, C.W.F.~Everitt, J.~Scargle, A.~Silbergleit and the members of the Gravity Probe~B theory group.
J.M. and Z.W. acknowledge the Fisher College of Science and Mathematics and Honors College at Towson University for travel support to present these results.

\end{document}